# An Intent of Collaboration: On Agencies between Designers and Emerging (Intelligent) Technologies

Lin, Pei-Ying[ab]*; Heij, Julie[ab]; Borst, Iris[ab]; Joosten, Britt[ab]; Andersen, Kristina[a]; IJsselsteijn, Wijnand[a]

[a] Eindhoven University of Technology, Eindhoven, The Netherlands
[b] Authors make equal contribution to the paper
* p.y.lin@tue.nl; pei@peiyinglin.net

Amidst the emergence of powerful intelligent technologies such as LLMs and text-to-image AIs that promise to enhance creative processes, designers face the challenges of remaining empowered and creative while working with these foreign digital partners. While generative AIs offer versatile, informative, and occasionally poetic outcomes, their lack of embodied knowledge presents an even greater challenge to designers in gaining fruitful outcomes, such as in the field of Digital Craftsmanship. In this project, three designers embarked on a three-month experimental journey with an intention to co-create with Google's LLM as a potential intelligent partner to investigate how it will influence the designers' creativity. We found that a power dynamic of agencies exists between the LLM and the designer, in which the designer can easily lose their creative agency. Regaining the designer's creative agency involves introspection into their own creative process, a structural understanding of the specific emerging technology involved, and deliberate adjustments to the dynamics of the human-technology relationship. We propose paying attention to the designer's inner world and parties of agencies when engaging with emerging intelligent technologies through three aspects: the sensitivity towards a creative process as cognitive activities; the active investigation into specific technology's capability; and the adjustment towards an appropriate working relationship between the designer and the emerging technology.

***Keywords: design agency; empowerment; more-than-human; creativity.***

## 1   Introduction

Designers are always at the frontier of adopting emerging technologies for wider design opportunities, especially in the field of HCI. Often, they are confronted with the challenge of adapting their creative processes to emerging technologies that were invented for an engineering interest instead of a designerly one. Where engineering pursues solutionist approaches focused on optimisation, perfection, and realism, design may favour exploratory approaches that celebrate glitches, embrace surprises, and generate novel aesthetics. Since the interest and problem space for engineering and design can be drastically different, these emerging technologies often lack considerations for design practices. The disparity between designers' needs within their creative processes versus the functionality of emerging technologies often demands that designers make active adjustments in their approaches, both process-wise and perspective-wise. In our research, we are interested in *finding the perspective adjustments that designers make that might bring innovation to the design practice itself*, similar to how the camera made artists think differently about imagery, compared to the paintings they knew when the camera was first invented. We took the textile making process as an exemplar for studying the meeting of physical (embodied) practices with digital intelligent technologies, which is often categorised as Digital Craftsmanship in HCI studies. We propose the textile making process as a complex process, where its complexity cannot be reduced to tackle the nuances within human and intelligent technologies collaboration (Lin & Witkowski, 2025) and position human cognition in the centre of study (Malafouris, 2013) for insights. Additionally, we use the gap between digital and the physically embodied for highlighting the mundane within the creative processes for observation (Crabtree et al., 2023).

Current emerging intelligent technologies are often designed in computer science labs and are being used in areas they were not designed for, such as the recent development of Artificial Intelligence (AI) since 2022 with Large Language Models (LLM) and text-to-image AIs. These platforms were invented through the exploration of algorithms and digital datasets that are based solely within the digital space. These emerging platforms tend to be powerful, with digital matters such as texts, concepts, and digital images, giving promises for making creative processes faster and more inspiring. However, when engaged with textile making, they became uninformative due to lacking the physical, tactile, and embodied information. Although concepts and images could work for the creative processes, they are often far from actionable or render the designers (and artists) non-creative (Halperin & Rosner, 2025). With the conflicting perception that these emerging generative AI platforms can be useful for creative practices but potentially also limiting and harmful, we decided to take a practice-based Research through Design (Frayling, 1994) approach to unpack and observe generative AI at work within creative processes. We asked how generative AI can be a designer's creative partner, and what it takes to work with AI.

In this project, we explore the nuance of creative processes both from the extrospective (the making process) and the introspective (designer's mental state) when designers are confronted with emerging technologies whose abilities were not tailored for their need but promised to be intelligent and potentially 'sentient' or 'autonomous'. Through taking AI as non-human within the More-Than-Human framework and setting up the challenge of having AI as a co-creation partner in textile making, this paper presents our journey working with LLM on textile making that was highly frustrating and disempowering and the effort of being empowered again. We approach the 'intent of collaboration' through the More-than-Human perspective and see it as an assemblage with distributed agency (Bennett, 2010; Wakkary, 2021) from the perspective of the humans, as opposed to the de-centring human approach (Nicenboim et al., 2024) to observe the natural changes designers would face introspectively. The analysis of our first-person perspective (Desjardins et al., 2021) working diaries was done by using frustration as the anchor to examine the emotional and cognitive changes in our making journeys. We have found that there exists a power dynamic between humans and AI, which requires designers to develop sensitivity to the positionality of AI, for the designer to regain their creative agency. Through this project, we have shown three mutually reinforcing aspects: 1) a demonstration of noticing the cognitive nuances within designers when working with emerging technologies that were designed without specific functional purposes; 2) the power play between humans and emerging technology within a creative process; 3) the dynamic of designer's diminishing creative agency and potential pathways for its reclamation. Our contribution suggests that working with emerging technology is a process of learning where agency plays an important role, and such agency is a dynamic, relational construct that requires deliberate considerations by the designer. To remain an empowered designer requires a good understanding of the cognitive activities within one's own creative practices and being conscious of the designer's own agency and the agencies delegated to the emerging technology they work with.

## 2 Background and Related Research

### 2.1 Phenomenal Emergence of Generative AI

In the spring of 2022, the launch of Midjourney (Midjourney, Inc., 2022), StableDiffusion (Stability AI Ltd, 2022) and a major update to DALL·E (OpenAI, L.L.C., 2022a) marked a turning point for the creative industry, as it catalysed the widespread adoption of generative AI tools in artistic and design practices. Prior to this, generative AIs such as Disco-Diffusion (alembics, 2022), adaptations of CLIP (Radford et al., 2021) and BERT (Devlin et al., 2019), or Google Deep Dream (Mordvintsev, 2015) had their presence exclusive to new media artists with a proficient level of programming abilities. This new generation of generative AIs runs on supercomputer servers where users can access them through the internet, returning impressive images within minutes by giving simple descriptive sentences as prompts. The following year, Large Language Models such as ChatGPT (OpenAI, L.L.C., 2022b), Claude (Anthropic, 2023), and Bard / Gemini (Google, 2023) made their debut to the public, demonstrating the ease of text generation for narrations, coding, prompt generations, conversations, and many more. These AI platforms demonstrate versatile usage for diverse tasks, where prompt engineering has become an art in itself to steer generative AI outcomes in the desired direction (Giray, 2023; Liu & Chilton, 2022; Meskó, 2023). It became an important question to look at how generative AI will influence the creative processes of designers.

### 2.2 HCI Research on Human-AI Creativity, Co-creation, and Collaboration

The use of AI for design in the field of HCI has been widely studied. In particular to creativity and creations within design processes, there are subjects such as AI creativity (Boden, 1998, 2004; Boden & Edmonds, 2019; Hsueh et al., 2024), AI-human collaboration and co-creation (Lin et al., 2024b; Liu et al., 2024; Rezwana & Maher, 2021), designer's expectation of AI (Yildirim et al., 2022, 2023), AI facilitated ideation (Davis et al., 2024; Jo et al., 2024; Lin et al., 2024b; Mahdavi Goloujeh et al., 2024; Shin et al., 2023), music generation (Park et al., 2024; Robson et al., 2024; Vear et al., 2024), programming (Ferdowsi et al., 2024), and searching for the design space (Fan et al., 2024; Huang et al., 2024; Shin et al., 2024). Whilst the popularity of generative AI raises concerns that designers struggle for their agency while using AI tools (Mügge, 2024) and that AI 'takes over all the fun part', making creative workers become non-creative labourers (Cremer et al., 2023; Mügge, 2024), the main trajectory of HCI study remains in the objective, observative position when looking at the human-AI relationship. But like the history of computational art has shown us that computers change the thinking and making of art, how generative AI will demand designers to change internally can also be expected. In this project, we shift our focus from the interaction between humans and AI to the proactive changes designers need in order to unleash their creativity while working with AI during the textile-making process.

## 2.3 Revealing the Nuances within Textile-Making Processes

Textile making has been known for being highly complex and dynamic, where materiality plays an important role in the process (Bennett, 2010; Ingold, 2010; Sennett, 2008). Textile making and other craft practices have been conceptualised as an assemblage with vibrancy (Bennett, 2010; Deleuze et al., 1988) that consists of humans, machines, and materials, where each entities have its own agency. This complex system is often observed and analysed through the More-than-human framework, which sees design as something the human and non-human machines and materials perform together (Wakkary, 2021) and which cannot be studied in simplification (Bennett, 2010).

In HCI research, Digital Craftsmanship emerges through combining digital systems and textile making, where human-computer interaction is being studied through textile making. For unpacking the complexity of textile making for human-computer interactions, documenting and investigating the details within the making process is highly valued (Goveia da Rocha et al., 2022a; Goveia Da Rocha et al., 2022; Lin et al., 2024a; Meiklejohn et al., 2024; Rutten et al., 2022). Methods of how to pay attention to the material, context, and landscape (Oogjes & Desjardins, 2024; Oogjes & Wakkary, 2022) have also been developed.

The delicacy of looking at the detailed interactions and happenings between humans, machines, materials, and AI makes the intersection of AI and digital fabrication a rich ground to investigate. For the purpose of observing the internal nuances of designers' creative processes, we set our challenge with an intent to co-create with AI in the highly embodied space of textile making. Our project is an attempt to dive deep into the perspectives of designers (humans) in this complex intersection to reveal the important nuances for designers to work in this space.

## 3 Methods

Our research investigates what is required for designers to work with emerging technology introspectively. We contextualised our research within textile making and generative AI, which allows us to adapt the tools of investigation developed for Digital Craftsmanship. In order to observe and analyse the textile-making processes in detail, a practice-based Research through Design approach is employed, which focuses on gaining knowledge through direct engagements in making rather than the final designed outcome (Frayling, 1994) which facilitate us to look at the dynamics of internal relationships between entities within the assemblage in situ.

The project was set as a semester-long group project with three master students (2nd, 3rd, and 4th authors) in Industrial Design in the autumn of 2023 as a given assignment. The authors took 2 months to create various textile samples with generative AI, where they documented the processes and had weekly meetings for reflection, and made adjustments accordingly.

The making process, presented as 'Studio Journeys' in this paper, goes as follows: We asked if we could 'co-create' with generative AI, in this case, the LLM platform Bard (Google, 2023). The making process began with each of the authors adopting one textile making technique (embroidery, crochet, and weaving), where they feed an image of trivial choice into Bard (Google, 2023), and ask for its contribution -- whether it be ideation, the technique of making, or reflection -- from which they would make the textile accordingly, document the process, and reflect. The authors had no prior expectation of what would be made from this process, and they expected to find out with Bard. This iterative image-making-reflection process had taken place twice with two different images, the first generated by Midjourney (Midjourney, Inc., 2023) (Figure 1) and the second by Canva Image Generator (Canva, 2023) (Figure 2).

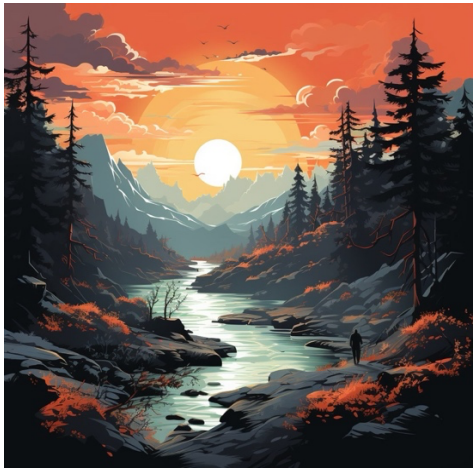

*Figure 1. Image of the sunset. Created by Midjourney.*

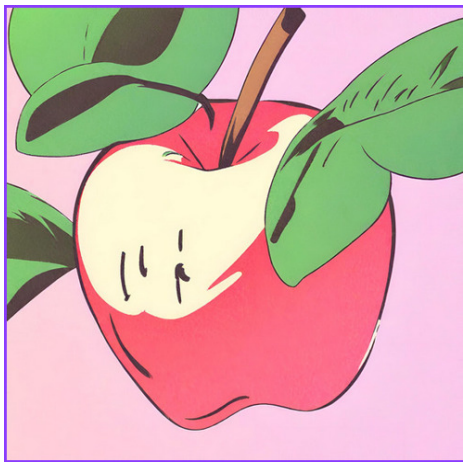

*Figure 2. Image of an apple created by Canva Image Generator.*

In the research, textile making was used as a method and the means to see the 'iterative journey as a research outcome' (Goveia da Rocha & Andersen, 2020). The process is documented by using a revised version of Goveia da Rocha's template for documenting samples (Goveia da Rocha et al., 2022a) with an additional section to keep track of the experiences from the conversations with Bard. The documentation is written in first-person perspective in order to explore the introspective aspect of the design process (Rutten et al., 2022; Tomico et al., 2012).

Three rounds of analysis and meta-analysis were done. The first analysis on the process and documentation was done immediately after the making process by the 2nd, 3rd, and 4th authors in a qualitative manner (Braun & Clarke, 2006). The second round of analysis and reflection was done 12 months after the making process, and the third round of analysis was done 15 months after the making process. The second and third rounds of analyses were done by the first author through semi-structured interviews and discussions with 2nd, 3rd, and 4th authors, along with meta-self-reflections from the 2nd, 3rd, and 4th authors.

## 4    Positionality of the Designers

In this project, the 2nd, 3rd, and 4th authors were engaged in the making process, who will be referred to as the designers. Each one of them has chosen to engage in one textile fabrication technique to work with Bard. The positionality of these authors is reflected in their unique journey with Bard; therefore is listed as follows.

The second author (2A hereafter) investigates our relationship with nonhumans from a first-person perspective through making. She aims to spark conversation with her designs about our possible futures and the role of the designer in this. The 2A chose to weave with Bard. She has no prior experience with weaving.

The third author (3A hereafter) is a student in Innovation Management and Industrial Design and works as a part-time artist. She designs and researches how design concepts can be put into practice and create impact. Her work aims to combine a strategic way of thinking with the unexpected outcomes of material engagements and creation. The 3A had chosen to embroider with Bard. She had no prior experience working with Embroidery.

The fourth author (4A hereafter) is a user-centred designer valuing aesthetics within her designs. In this research, she uses the fabrication method of crochet when co-creating with Bard. She has two years of experience with crochet, making clothes and accessories for personal projects.

## 5    The studio journeys

Three designers' attempts at co-creation with Bard are presented in the form of 'studio journeys' composed of textile samples following the textile-making development timeline in Figures 4, 5, and 6. We use the notion of 'studio journeys' in reference to 'becoming travellers' in open-ended material making proposed by Goveia de Rocha and Andersen (Goveia da Rocha & Andersen, 2020). The figures are to capture the internal cognitive struggles and shifts in correlation to the process of making. The detailed interaction between the designers and Bard is further presented with selected quotes from the conversation records with Bard. The journeys are divided into sub-journeys according to each sample. The designers supplement each sub-journey by scoring their agency versus Bard's agency to emphasise the fluctuation of agency between the designers and Bard.

### 5.1    'Agency' – the parameter that came in the third round of analysis

During the second round and third round of analysis, we introduced parameters as tools to facilitate meta-reflections on our qualitative analysis. The parameters were not used for traditional quantitative evaluation, but as anchor points to initiate discussions on the focus themes in-depth. For instance, in the second round of analysis, parameters 'Acceptance', 'Collaboration', and 'Frustration' were used as a guided discussion. These parameters later generate the discussion on 'agency'. The evaluation of 'agency' took a long discussion among the authors in the process of trying to identify what had happened between the designers and Bard.

The designers noticed themselves being extremely frustrated and disempowered through the interaction with Bard. During the first analysis that took place immediately after the making process, the designers took the phenomenological outcome as strategies that the designers developed to encounter their frustrating experience working with Bard. However, the designers find that strategies do not speak for the main emotional and cognitive effort required to make their co-creation with Bard fruitful. A second attempt to depict the emotional and cognitive effort was made by introducing the scoring of 'Acceptance', 'Collaboration', and 'Frustration'. 'Acceptance' was used to describe each designer's mentality towards accepting Bard's capabilities in relation to their expectation of Bard during each textile-making process. The designers assumed Bard could be powerful and creative, considering many people were generating images, prompts for images, prose, or even programming codes with LLMs during that time (autumn of 2023). They slowly figured out that Bard was not how they expected it to be, not less powerful, but with different capabilities. 'Collaboration' was an emotional-cognitive parameter to evaluate the actual interaction versus the expected interaction of 'collaboration' with Bard as a partner on equal footing. 'Frustration' was the parameter to trace the emotional activities of the designers in the making process. The authors find 'Acceptance', 'Collaboration', and 'Frustration' are still insufficient to describe both the internal emotional and cognitive understanding and active adjustments of the designers, nor do they describe the dynamic nature of Bard as a multifunctional LLM.

The rating of 'Agency' of the two parties, the designer(s) and Bard, was eventually used to *describe the dynamic negotiation of agencies between the human and the LLM as a perceptual description from a first-person perspective, the Sense of Agency (Gallagher, 2012)*. For which, the 'agency' influences the designer's decision. The 'Designer's agency' describes how much agency the designer felt to have during that period of time, whereas 'Bard's agency' describes how much agency has been given to Bard in the perception of the designer. These ratings were evaluated by each designer subjectively, in the argument that the agencies of the two parties were the result of the designer's own cognitive construct. The ratings were not exact scores but vague ranges, where Bard's agency and Designer's agency should be read together. As evidenced by the designers reflections, Bard's position is constantly renegotiated: sometimes in a dominant position, sometimes in a subordinate position, sometimes as a technician, and sometimes as nobody. These cognitive positions for Bard highly influence their subjective perception of creativity and agency, which comes as waves of disempowerment and empowerment as shown in Figures X. This perception of creativity and agency is fully internal, organic, and does not reflect their satisfaction of the final textile outcome.

## 5.2 Plotting the journeys

The journeys are plotted in the second round of analysis and revised in the third round of analysis. They were the tools that helped to distil the cognitive and emotional activities taking place within the designers, both through quotes and textile samples, in order to allow textile samples (external results) to be read in parallel to first-person reflections (internal activities). The journeys can be read following the description of Figure 3. The timeline in the centre is constructed chronologically according to each designer's conversation with Bard. Each colour of the timeline is one 'chatroom' with Bard. Bard works in a manner that users can open a chatroom with Bard, where Bard will have memories of all the previous conversations held within the chat room. Starting a new chat room means getting rid of previous memories and beginning a new conversation without memories. In the context of our project, the designers can make multiple samples while staying in the same chat room or start a new chat room. We divide the making processes according to each designer's definition of 'stage', which is similar to a subdivision of a project, indicating the mental stop point for each concept. The stages are indicated as shaped backgrounds with gaps in between. The quote selected for each stage shown is the designer's summary of their interaction with Bard. The satisfaction score is the designer's satisfaction towards the final textile outcome shown in the image. The turning point marks an important cognitive shift that the designer realised in their second round of analysis, which made a difference in how they feel towards their interaction with Bard.

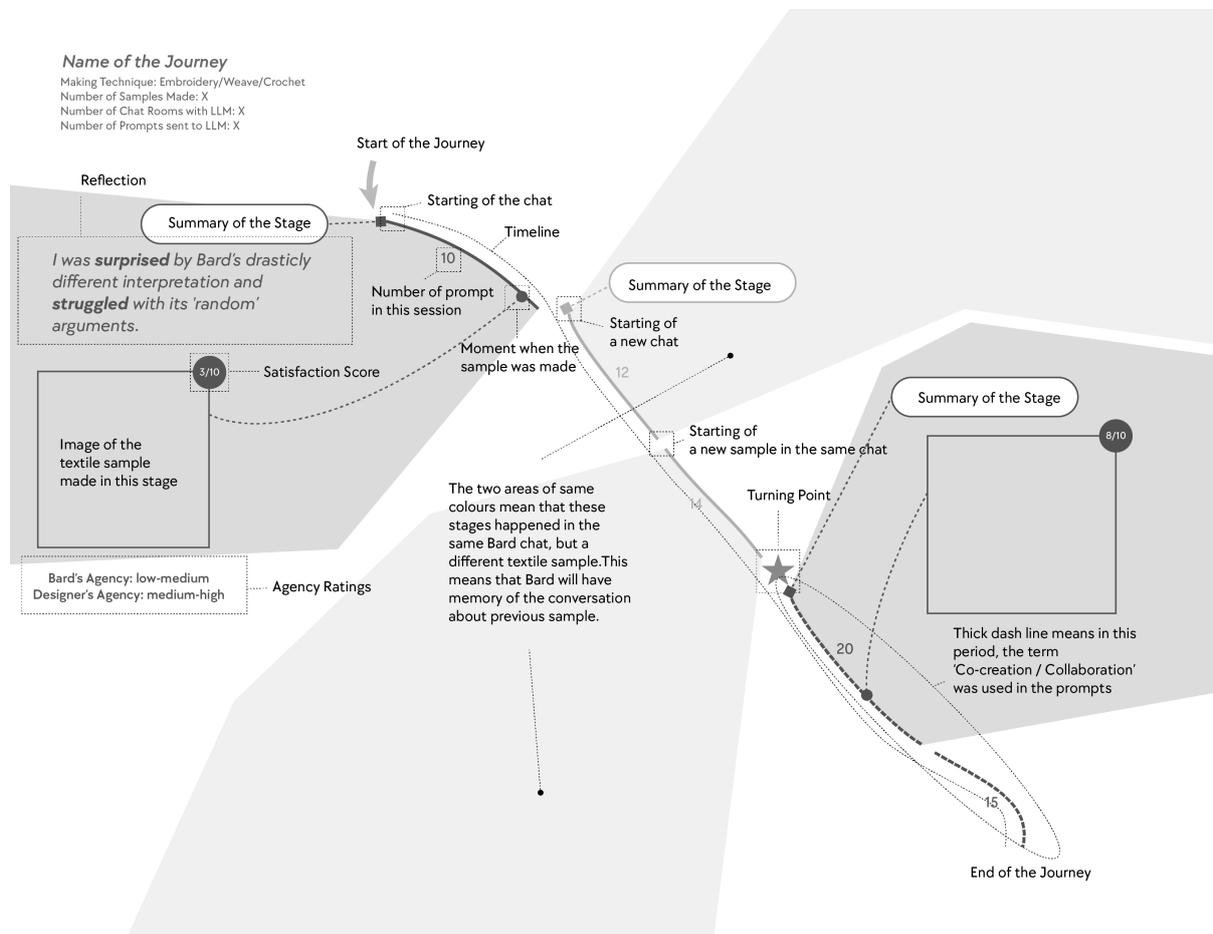

*Figure 3. Explanation of how to read the journeys.*

### 5.2.1 The embroidery journey

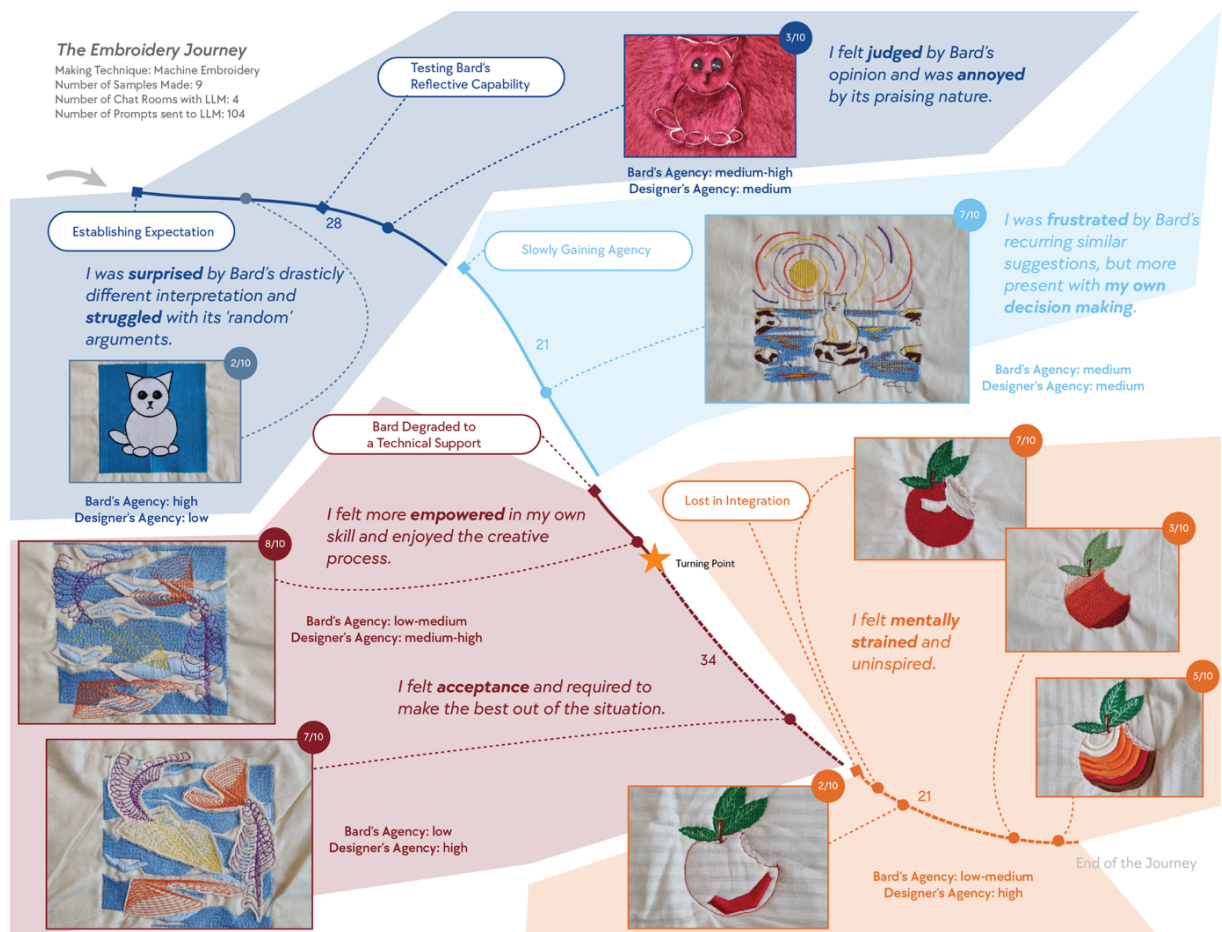

*Figure 4. Studio journey of the 3rd author using embroidery as the technique.*

The journey of the third author (Figure 4) consists of four different Bard conversations marked by four different coloured backgrounds. The third author's journey began with trying to understand Bard's ability and putting themselves in a secondary position. 'I decided to follow Bard in its chain of thoughts and was surprised by its opinionated nature.' The third author mentioned that Bard had mistaken the sunset image (Figure 1) for a cat. They followed Bard's instruction and made the embroidered cat, took a photograph and uploaded it to Bard for its opinion. Bard criticised the cat and said everything was 'a mismatch'. The third author decided to still follow the cat direction but have more personal input by choosing a hairy fabric for embroidery, which they uploaded to Bard and received praise for. The third author became frustrated while making the third sample because Bard began to have recurring answers. The frustration led the third author to take charge of creative decisions and position Bard merely as a technical support in the third stage. During this stage, the third author felt empowered, while Bard's agency declined. The turning point marks a significant cognitive realisation for the third author to consciously know where to position Bard in their making process. Immediately after the turning point, we see that the Designer's agency became higher than Bard's agency, and the author is also more satisfied with the process. The enjoyment disappears in the last conversation due to recurring responses that 'Bard started providing the same suggestions over and over again', although the Designer's agency is still higher than Bard's.

The dashed line marks the period where 'collaboration' was actively used in the prompts. However, we noticed that collaboration as a prompt does not guarantee to be truly 'collaborative', which is a recurring phenomenon in all three journeys where the designers became unmotivated by the end.

### 5.2.2  The crochet journey

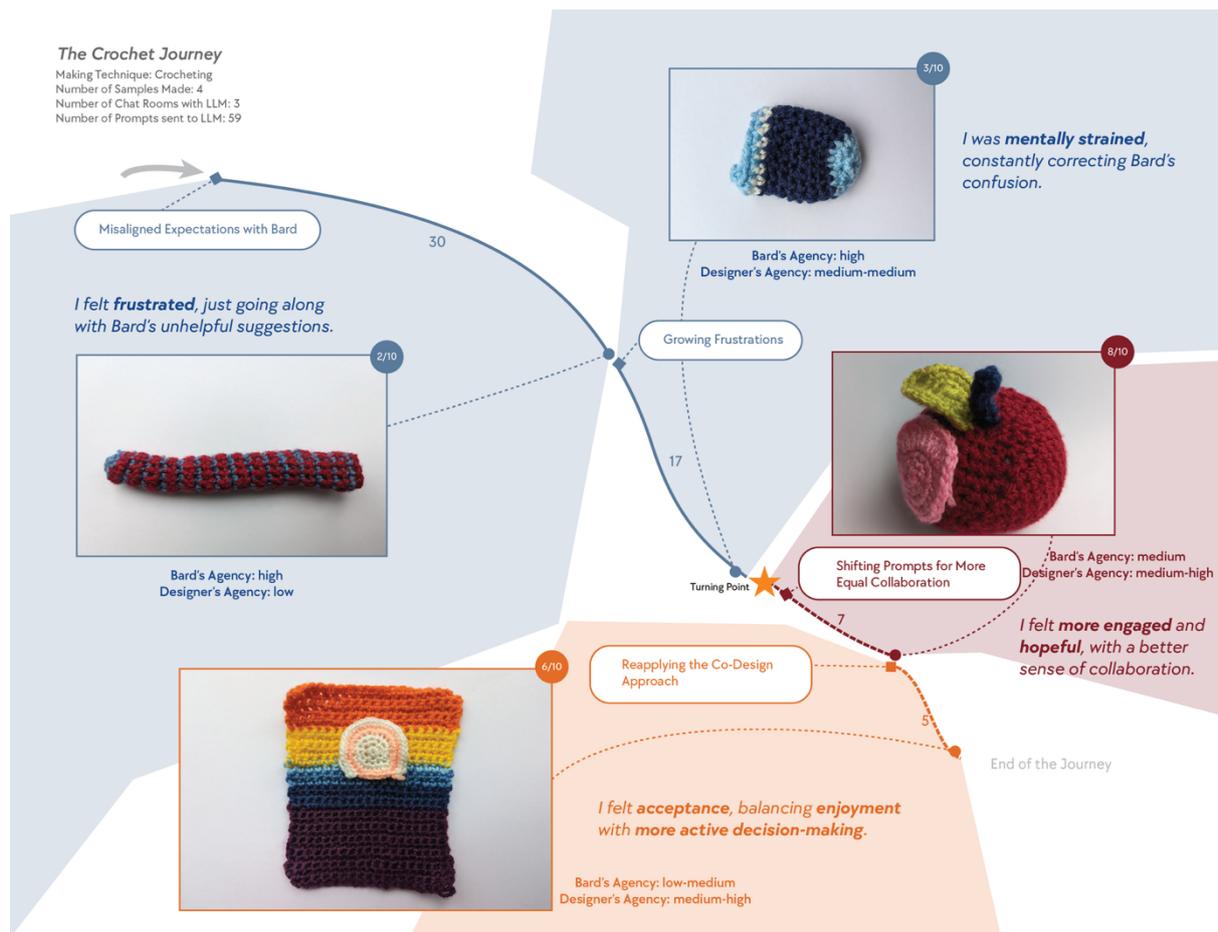

*Figure 5. Studio journey of the 4th author using crochet as the technique.*

The fourth author's journey consists of three chats with Bard, divided into four stages (Figure 5). The fourth author felt frustrated upfront due to Bard's unhelpful suggestions that 'Bard could not generate a crochet pattern.' The fourth author experimented to give Bard full agency: 'I mostly followed Bard's suggestions, hoping they would eventually lead somewhere useful. Bard also began interpreting the image differently after a while (as a water droplet).' The fourth author made the water droplet, but also decided to tell Bard that it was 'wrong', which sets the moment that the fourth author proactively took charge of the process. It was also the moment of a turning point. The image of an apple (Figure 2) was uploaded to Bard, where the fourth author asked Bard to 'co-design with me to make a crochet artwork based on the image'. During the third stage, the Designer's agency became slightly higher than Bard's agency. The author also felt much more engaged in the process, where 'it became more of a discussion than blindly following a plan.' For the fourth stage, the fourth author wrote that 'the process overall was much more enjoyable because of the acceptance of Bards' abilities, his strengths and weaknesses. This awareness made me more engaged in decision-making, which ironically made the process feel slightly less collaborative than before. However, the overall experience was far more satisfying than in earlier samples as it showed how framing the prompt could influence the creative dynamic.' In which the human-LLM relationship became less of a collaboration, but the designer felt more creative.

### 5.2.3   The weaving journey

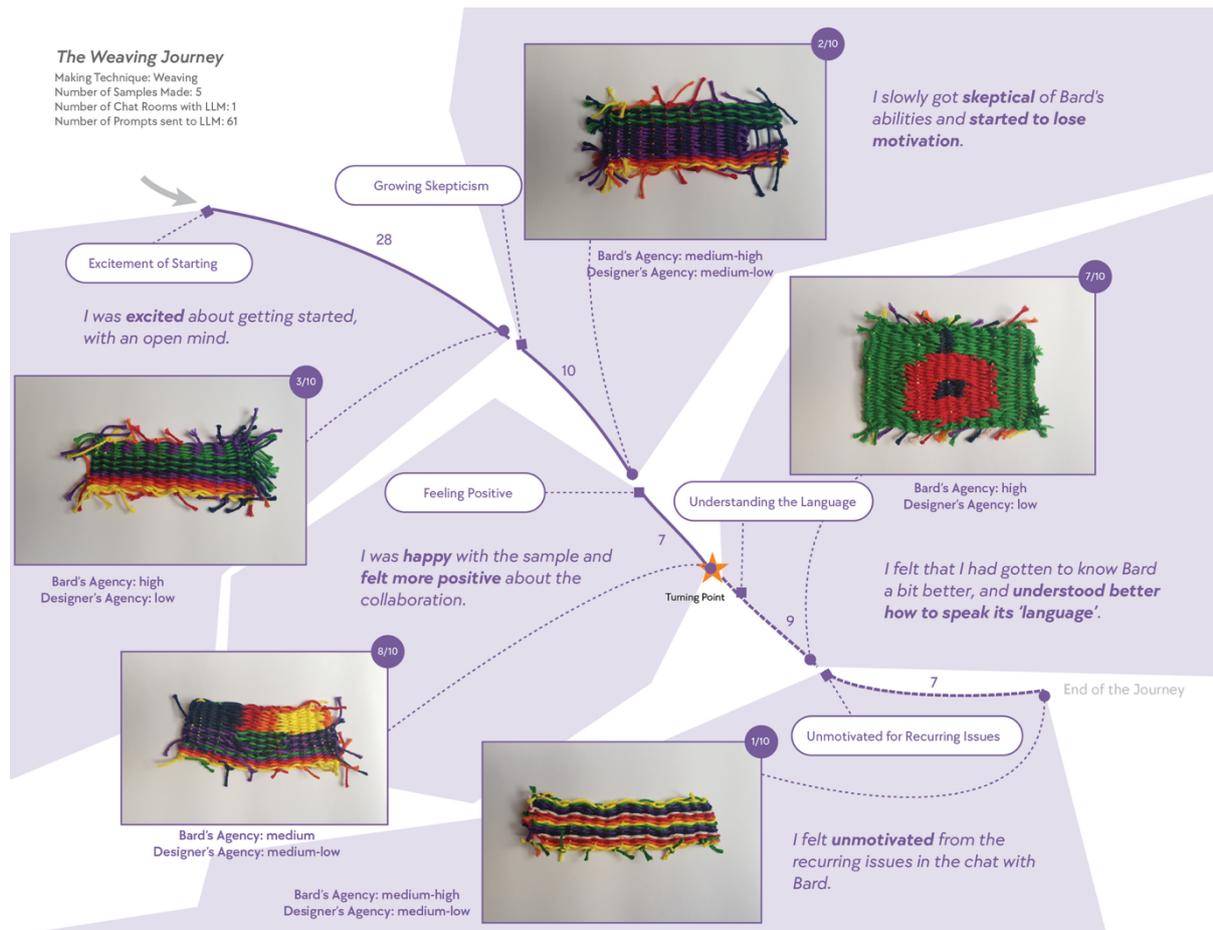

*Figure 6. Studio journey of the 2nd author using weaving as the technique.*

dhe second author's journey consists of one continuous chat with Bard divided into five stages. The journey began with excitement. From Figure 6, we see that Bard's agency was high while Designer's agency was low during the first stage. The 2nd author wrote in their note in the second reflection: 'I had the expectation that Bard would be able to generate patterns that would be easy to follow, and that Bard would not give multiple options for me to choose.' And her 3rd reflection: 'I felt dissatisfied with the level of the pattern that Bard had generated for me. …. I decided to make a simple pattern that Bard suggested, which is a very simplified version of the input image.' In the second and third samples, we see the designer's agency grew while Bard's agency declined. It was triggered by Bard's incapability for precision: 'It was in this part of the chat that I started to notice that Bard's "memory" was very limited. I had to repeat the colours I had available for weaving multiple times, which made the conversation rather frustrating.'

During the whole journey, the second author always delegated higher agency to Bard and never took control themselves. The Designer's Agency is always rated lower than Bard's agency on the journey. The second author expressed that this strategy was taken for the purpose of truly wanting to collaborate with Bard, which still failed in the end due to Bard being 'repetitive' and not giving new insights. However, the moment when the second author felt that 'I had gotten to know Bard a bit better and understood better how to speak its "language"', they explained, came after they gained awareness of what they wanted as a designer. Such awareness is gained through fully understanding the exact cognitive nuances engaged within the creative process and being capable of being explicit about it while communicating with LLM.

## 5.3    Four phases of power dynamics between the designer and the LLM

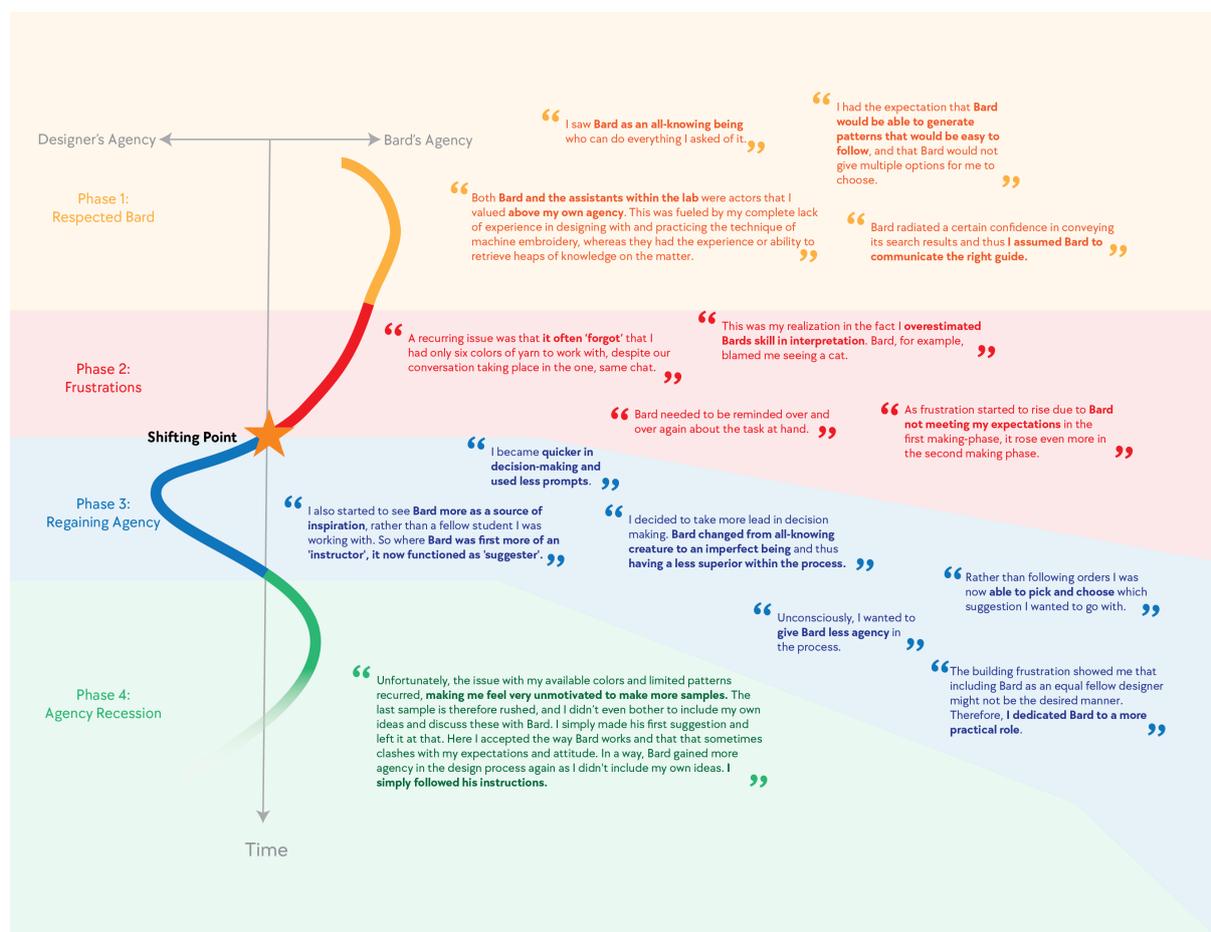

*Figure 7. Quotes from the self-reflection of the designers showing how the agency fluctuates between the designer and the LLM internally.*

We noticed that there were 4 phases of development (Figure 7) where the definition of collaboration and agency fluctuates. In the first phase, Bard is being treated as the collaborator who knows all, indicating that the designers delegated high agency to Bard. In this phase, the designers entered prompts in Bard, received instructions from Bard, and executed accordingly. The designers prioritise Bard before themselves. In the second phase, the designers started to notice the incapabilities of Bard, such as being forgetful, repetitive, and misinterpreting images. The designers went through the process of improvising the prompts and trying to make Bard work in the way they anticipated. Unfortunately, Bard still could not meet the expectation.

In the third phase, two designers (3A and 4A) decided to regain their own agency by having the confidence to take the lead, telling Bard that its answers were wrong. They clearly addressed the task Bard will need to perform for them and emphasised the word 'collaboration' in the chat. The designers treat Bard's answers as inspirations and suggestions and see Bard as an assistant. As a result, the designers observed that there are fewer prompts in these stages because they follow their own plans rather than Bard's. They also find this stage much more comfortable as a creative creation session. For 2A, although they did not proactively regain their own agency with Bard, they also noticed that being explicit about what they wanted from Bard made the collaboration feel much more comfortable. For all the designers, they went through a learning curve of understanding Bard's capabilities as well as a detailed understanding of the making process of their chosen textile technique, which allowed them to become empowered to make design decisions and find a working communication method. In the fourth phase, a recession of agency took place. The designers got tired and let Bard take the lead. However, this also made the designers more frustrated.

We see that because the designers were working with a novel and powerful technology, most of their effort within the creative process went into understanding the technology and navigating through the struggle of agencies. In which, the process made them learn about the nuances of their cognitive activities required for the creative textile making processes and acquire a conscious working positionality with the technology.

## 5.4    Agencies and mechanisms of regaining agencies

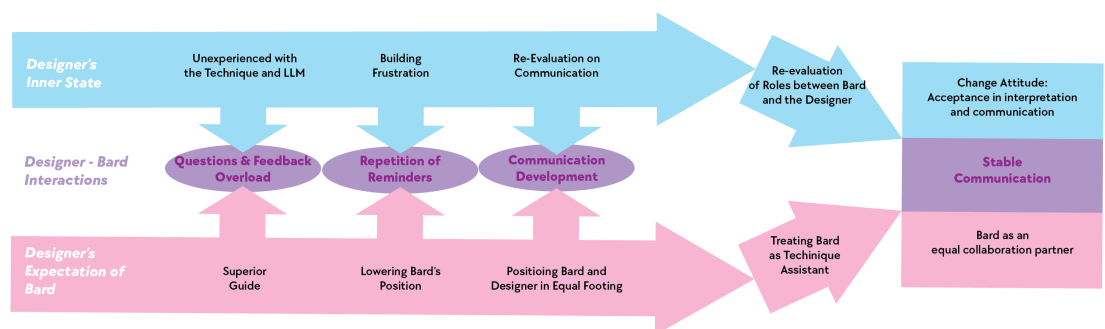

*Figure 8. The mechanisms of relational agencies.*

We have observed that the power play between the designer and LLM is relational and dynamic, and we dissect the mechanisms in Figure 8 in our interaction with Bard. The power dynamic plays between the designer's internal status and the designer's expectation of the LLM they work with, where the designer-LLM interaction provides opportunities for understanding and adjustments. In Figure 8, one can see that the designer's expectation of Bard implicitly indicates the level of agency Bard has been given. In other words, Bard did not voluntarily choose a level of agency itself, its agency was delegated by the designer. The power dynamic between the designers and Bard was an organic process that consisted of four stages of interactions: 1) Questions & Feedback Overload, 2) Repetition of Reminders, 3) Communication Development, and 4) Stable Communication. Each stage of interaction is a phenomenological manifestation of the designer's expectations and internal condition. In the first stage, questions and feedback were overloaded. The designers were overwhelmed by the number of questions asked and the amount of feedback given. It happens under the premise that the designer is expecting Bard as a superior guide, a wizard-that-knows-all. In this stage, the designers did not acknowledge their own expertise and let Bard take control. In stage 2, frustration towards Bard built up for the designers. By repetitively repeating their prompts to Bard, the designers also slowly realised that a superior guide is not the right perception for Bard. This results in lowering the relative positioning and agency for Bard. At the same time, the designers also slowly gain their own agency by adjusting the positioning of themselves. The designers then slowly allow their own existing knowledge of design to work alongside AI. In stage 3, a turning point is reached. The designers gave up their expectations of Bard and reckoned that some mental adjustment was needed. They re-evaluate their communication technique, such as using 'co-design' and 'collaboration' in the prompt and expect less from Bard, which puts the designer and Bard on an equal footing. In stage 4, the designers obtained a reasonable expectation of Bard that fits with its capabilities, and the communication went smoother than before. The designers also reckon that at this stage, they finally felt that they could work with Bard constructively.

# 6  Discussion

## 6.1  Importance of self-reflection while working with emerging technology

Detailed self-reflection and documentation enabled us to observe the designer's inner world when working with emerging technologies in highly complex and dynamic processes, and notice the power dynamic of agencies between humans and technology. In this project, we began by asking how to 'co-create' with a LLM for an open-ended exploration. LLMs are an emerging technology in the sense that their abilities are versatile, rapidly updated, and there are no available protocols to work with. This entails that a certain level of learning and exploration is needed for humans while incorporating it into their own making processes. For designers, to have a fruitful 'co-creation in an open-ended exploration' with any emerging technologies is essentially a question of how the designer can feel 'creative' while working with emerging technologies. This topic is subjective and highly complex, for too many aspects are involved, such as materials, machines, and context. In section 5.2, we see that it required three rounds of analysis and meta-reflections to identify that it was the agency dynamic that contributed to the designer's frustration working with Bard. By focusing on the documentation and reflection of the process attached to an outcome instead of only evaluating the outcome, we noticed that satisfaction with the process does not relate to satisfaction with the outcome. This raises questions about the popular HCI research approach, where the generative AI outcomes were evaluated for creativity by the participants without asking for their reflections on the process. Existing research such as Portfolio of Loose Ends (Goveia da Rocha et al., 2022b) captures the dynamic of influence between concurrent projects within the same space, and Design Bookkeeping (Meiklejohn et al., 2024) which records design process, samples, and decisions as a ledger, we argue that introspective reflections can reveal nuances of human-technology interaction to allow us find ways to work with emerging technologies.

## 6.2 'Agency' of the more-than-human partner is a relational concept that requires deliberate decisions from the designer at work

Section 5.5 shows that the perceived 'agency' of Bard by the designers is tightly related to the designer's expectation of Bard. Bard, as the non-human collaborator in the three studio journeys, is positioned with different roles due to different expectations from the designer. The designer perceives their own agency versus Bard's agency according to how they see their own role and position within the collaboration relationship to Bard's given role, and this changed throughout the multi-stage engagement. Designers working with tools rarely have the need to think about what they expect from the tool, nor what agency should be given to the tool. However, working with more-than-human partners whose capabilities are implicit will bring designers into the space where expectations can be versatile, modes of interactions can change, and a designer's inner status fluctuates, resulting in what a designer might feel as a power dynamic between themselves and the more-than-human partner. In order to be empowered in the process of working with more-than-human partners, we argue it is important to be conscious of the relational agency between the designer themselves and the more-than-human partner, and to make decisions accordingly.

In *Things We Could Design*, Ron Wakkary argued on the basis of vital matters, that things have agentic qualities that 'contribute to their own making and the making of other things', which further led to the discussion of distributed agency within the assemblage that consists of all the entities engaged within the making process (Wakkary, 2021). How the agencies are distributed within the assemblage can be case-specific to the composition and dynamic of the assemblage, for instance, different when it is designed with a plant that has biological needs versus when it is designed with a computer algorithm that does not have biological needs (Montemayor, 2023). In our specific case, Bard has participated in the assemblage as an LLM without its own intention, where its agency was delegated by the designer through the designer's expectation of it. This relational nature of human-technology relationship reveals an important aspect: Whilst whether AI can have 'agency' and 'autonomy' is still debatable within the field of Artificial Intelligence, since it is highly dependent on how each AI is being designed and coded, as well as depending on the philosophical level it is being discussed - in practice, we noticed that the human's perceived agency of the AI is relational. Such a relational nature is correlated with the complexity of AI, which appears to be versatile and multi-functional, like a powerful black box. This relational agency should also be studied when looking at creativity agencies in human-technology collaborations.

# 7  Conclusion

We used Bard as an exemplar of emerging intelligent technology that is powerful, versatile, but not tailored to our textile-making process, to explore the challenges and struggles designers might face. Through in-depth documentation and self-reflection of our journey working with Bard, we found that designer's sensitivity towards their inner positionality is crucial in navigating collaborations with technology. The notion of felt 'agency' plays an important role in feeling creative and empowered within a studio making process. We also notice that the felt agencies of the designer and of Bard are relational and dynamic, and that the process of making involves a process of the designer adjusting their expectation and interactions with the new, unknown technology they work with. We propose that in this time of Artificial Intelligence bloom, where technologies become more sophisticated and powerful, paying attention to the relational agencies between the designer and the technology by looking introspectively is key for creative professionals to remain creative and empowered. Our research also highlights the importance of thinking about the positionality of the designer ourselves when working in the more-than-human world, where Research-through-Design working methods with detailed documentation and self-reflections can provide keys for future design education in times of close human-technology collaboration (Khosravy et al., 2024). Due to the timeframe in which this project was conducted, we did not have the opportunity to explore more powerful LLMs and generative AI that would potentially demonstrate much more sophisticated behaviour and abilities. Additionally, emerging technologies such as nanotechnology, biotechnology, and augmented reality lie beyond the scope of the current study. However, their potential impact on designers' inner worlds presents a compelling avenue for future exploration. These areas will be left for subsequent research to investigate the evolving complexity of human–technology interaction and collaboration within creative processes.

**About the Authors:**

**Pei-Ying Lin:** (http://peiyinglin.net) is an artist, designer, and researcher in Science/Art who works with More-than-Human others including viruses, bacteria, textile, and AI. She is a PhD student at TU/e, resident artist of RepliFate, F Fellow of Royal Shakespeare Company and MIT ODL.

**Julie Heij:** is a master student in Industrial Design at TU Eindhoven. Her research explores relationality and slowness in design through making. She investigates how textile tools and techniques can challenge acceleration while fostering responsibility, attunement, and more sustainable, interconnected practices.

**Iris Borst:** is a master's student in Innovation Management and Industrial Design at the TU Eindhoven. She focuses on system-oriented design with an emphasis on multi-stakeholder inclusion and interdisciplinarity. Her work aims to bridge academic insights with practical applications to support industry transitions.

**Britt Joosten:** is a Master graduate in Industrial Design at TU Eindhoven. Her design research explores mental wellbeing and user engagement, with a focus on embodied interaction. She values aesthetics, usability, and simplicity in tangible designs that enhance daily life.

**Kristina Andersen:** is an associate professor at the Department of Industrial Design at Eindhoven University of Technology (TU/e). Her work is concerned with how we can allow each other to imagine our possible technological futures through digital craftsmanship and material practices.

**Wijnand IJsselsteijn:** Prof. Wijnand IJsselsteijn is a psychologist studying how technologies like VR and AI shape human behaviour, communication, learning, and wellbeing. His recent work emphasises human-robot interaction, AI in dementia care, and technology ethics. He holds the 2024/25 Distinguished NIAS-Lorentz Fellowship.

**Acknowledgement:** This research is made possible by EAISI, the Eindhoven Artificial Intelligence Systems Institute, through the Exploratory Multidisciplinary AI Research (EMDAIR) grant (Project no. 64). We would like to also thank Stephan Wensveen and Lenneke Kuijer for providing the opportunity to conduct this research within their Constructive Design Research course at Eindhoven University of Technology, and the technical supports from Wearable Senses Lab at Industrial Design Department at Eindhoven University of Technology.